\documentclass[letterpaper,10pt,conference,dvipsnames]{ieeeconf} 
\let\labelindent\relax
\IEEEoverridecommandlockouts                          
\usepackage{cite}
\usepackage{enumitem}
\usepackage{amsmath,amssymb,amsfonts}
\usepackage{algorithmic}
\usepackage{graphicx}
\usepackage{textcomp}
\usepackage{xcolor}
\usepackage{theorem}
\usepackage[font=footnotesize,skip=0.1pt]{caption}
\usepackage{tikz}
\usetikzlibrary{shapes.geometric,decorations.pathreplacing}
\usepackage{pgfplots}
\usepackage{pgf}
\definecolor{MyRed}{HTML}{e6550d}
\definecolor{EPFLRed}{HTML}{b51f1f}
\definecolor{MyBlue}{rgb}{0.20, 0.6, 0.78}
\definecolor{MyGreen}{rgb}{0.4,0.8,0.4}

\newcommand{\change}[1]{\textcolor{black}{#1}}

\DeclareMathOperator{\R}{\mathbb{R}}
\DeclareMathOperator{\Z}{\mathbb{Z}}

\newtheorem{theorem}{Theorem}
\newtheorem{lemma}{Lemma}
\newtheorem{remark}{Remark}

\allowdisplaybreaks[4]

\begin{document}

\title{\LARGE \bf Robust Resource-Aware Self-triggered Model Predictive Control}

\author{Yingzhao Lian, Yuning Jiang, Naomi Stricker, Lothar Thiele, Colin N. Jones
\thanks{This work has received support from the Swiss National Science Foundation under the RISK project (Risk Aware Data-Driven Demand Response), grant number 200021 175627, and under the NCCR Automation project, grant agreement 51NF40\_180545. (Corresponding author: Yuning Jiang)}
\thanks{Yingzhao Lian, Yuning Jiang and Colin N. Jones are with Automatic Control Laboratory, EPFL, Switzerland.{\tt$\{$yingzhao.lian, yuning.jiang,colin.jones$\}$@epfl.ch}}
\thanks{Naomi Stricker and Lothar Thiele are with Computer Engineering and Networks Laboratory, ETH Zürich, Switzerland.{\tt $\{$nstricker, thiele$\}$@ethz.ch}}}

\maketitle

\begin{abstract}
The wide adoption of wireless devices in the Internet of Things
requires controllers that are able to operate with limited resources,
such as battery life. Operating these devices robustly in an uncertain
environment, while managing available resources, increases the
difficultly of controller design. This paper proposes a robust
self-triggered model predictive control approach to optimize a control
objective while managing resource consumption. In particular, a novel
zero-order-hold aperiodic discrete-time feedback control law is
developed to ensure robust constraint satisfaction for continuous-time
linear systems.
\end{abstract}

\section{Introduction}
The operation of devices in the Internet of Things (IoT) networks and wireless sensing systems are strongly impacted by resource factors, including battery life and hardware longevity. In order to avoid unnecessary resource consumption caused by extra device triggers/updates, \change{such as the cold-boot power of a sensor,} the  controllers with aperiodic triggers can be deployed under self-triggered and event-triggered control schemes~\cite{heemels2012introduction,wildhagen2020resource,Lian2020,Lian2021}. In particular, the control action under an event-triggered scheme is updated \emph{reactively} by monitoring a trigger condition, whereas a self-triggered scheme updates \emph{proactively} by planning the next trigger instant in advance, leaving sensors and controllers in idle mode. Due to the limitation of the resource factors, especially battery life, a self-triggered scheme can be preferable and is, therefore, the research object of this work. \change{More applications refer to~\cite{stricker2021distributed}~\cite[Section 4]{wildhagen2020resource}.}

On top of the decision of triggering time sequences, the operation of a self-triggered device in an uncertain environment requires extra consideration of uncertainty propagation due to the lack of state measurement between two consecutive triggers. Most works decouple the triggering time decision from uncertainty propagation. For example, a tube-based method~\cite{aydiner2015robust,brunner2016robust} and a Lipschitz constant based method~\cite{li2014event} have been applied to quantify the uncertainty evolution. In discrete-time systems, the triggering time sequence has been chosen to maximize the duration of open-loop operation~\cite{brunner2016robust} or by monitoring the discrepancy between nominal performance and actual performance~\cite{dai2020fast}. Similar strategies have been applied to continuous time systems, where either the system is discretized~\cite{farina2012tube} or the actual state is compared to a nominal state with continuous state measurements~\cite{li2014event}. 

In this work, we consider a resource-aware self-triggered control problem for an uncertain continuous-time linear system, \change{where, to the best of our knowledge, no exisiting results can be directly applied in a numerical reliable way.} Notice that because the system is confined by limited resource factors such as battery life, continuous state measurement is impractical. Instead, we unify the triggering time sequence decision\change{, feedback gain} and the control input selection within one optimization problem. The main contributions of this work are summarized as follows:
\begin{itemize}
    \item \change{a novel decomposition of the dynamics into those linked to process noise and those to feedback dynamics is proposed. Accordingly, we present the \textbf{continuous-time} ellipsoidal set propagation dynamics driven by a \textbf{discret-time} affine feedback control;
    \item a robust resource-aware self-triggered MPC scheme is proposed that enables a unified decision of triggering time sequence and robust control input sequence;
    \item numerically stable implementation details are provided.}
\end{itemize}

\textbf{Outline} \change{Even though this paper study uncertain linear dynamics, to better introduce the concept, Section~\ref{sect:pre} starts with reviewing deterministic resource-aware model predictive control in a generic form, alongside} differential inequalities and some necessary results from ellipsoidal calculus. The main results are elaborated in Section~\ref{sect:main}, where dynamics driven by the proposed control law are summarized in Lemma~\ref{lem:ellip_dyn}. A numerical validation of the proposed controller is given in~\ref{sect:num}, followed by a conclusion in Section~\ref{sect:conclusion}.

\textbf{Notation} The Minkowski sum of two sets $\mathbb X,\mathbb Y\subset \mathbb{R}^n$ is denoted by $\mathbb X\oplus \mathbb Y = \{x+y \mid x\in \mathbb X,\, y\in \mathbb Y\}$. The set of symmetric positive (semi-)definite matrices in $\mathbb{R}^{n\times n}$ is denoted by $(\mathbb{S}^n_+)\mathbb{S}^n_{++}$. An ellipsoid in $\R^n$ centered at $q\in \R^n$ is defined as $\mathcal{E}(q,Q):=\{q+Q^\frac{1}{2}v\mid v^\top v\leq 1\}$ with $Q\in\mathbb{S}^n_{++}$. The support function of a convex set $X\in\R^n$ is defined by $V[X](c):=\max_z\{c^\top x\vert x\in X\}$ for all $c\in\R^n$. The notation $\Z_a^b=\{z\in \mathbb{Z} \mid a \leq z\leq  b\}$ is used to denote integer ranges and we use the notation $\mathbf{0}$ to denote the zero matrix. The set of compact subsets of $R^{n}$ is denoted by $\mathbb{K}^n$, and the subset of compact convex subsets $\mathbb{K}^n$ by $\mathbb{K}^n_\mathrm{C}$.

\section{Preliminary}
\label{sect:pre}

\subsection{Deterministic Resource-aware Model Predictive Control}
The dynamic of a linear time invariant (LTI) system in continuous time is given by 
\begin{align}\label{eq::dyn}
   \forall t\in[0,\infty),\;\frac{dx(t)}{dt} = Ax(t)+Bu(t)
\end{align}
with coefficient matrices $A\in\mathbb{R}^{n_x\times n_x}$ and $B\in\mathbb{R}^{n_x\times n_u}$, state $x(\cdot):[0,\infty)\to\mathbb{R}^{n_x}$, control input $u(\cdot):[0,\infty)\in\mathbb{R}^{n_u}$. Here, the state and control inputs are subject to constraints 
\[
\forall t\in[0,\infty),\;x(t)\in\mathcal{X},\;u(t)\in\mathcal{U}
\]
with constraint sets $\mathcal{X}\subseteq\mathbb{R}^{n_x}$ and $\mathcal{U}\subseteq\mathbb{R}^{n_u}$. 

In the context of self-triggered control scheme, the control inputs are changed at triggering time instances $\{t_k\}_{k=0}^{N-1}$. Therefore, one can represent the zero-hold control inputs by using the direct optimal control approach~\cite{bock1984multiple} over the time horizon $[0,t_N]$, i.e., 
\begin{align}\label{eqn:deter_ctrl_1}
    u(t) = \sum\limits_{k=0}^{N-1}v_k\cdot\zeta_k(t,t_k,t_{k+1}),
\end{align}
with $v_k\in\mathbb{R}^{n_u}$ the coefficients, and $\zeta_k\in\mathcal{L}^2[t_0,t_N],\;k\in\Z_0^{N-1}$ model the triggering property with a piece-wise constant function
\begin{align}\label{eqn:deter_ctrl_2}
    \zeta_k(t,t_k,t_{k+1})=\begin{cases} 1 & t\in(t_k,t_{k+1}]\\
    0& \text{otherwise}.
    \end{cases}
\end{align}

The update of the control input at each triggering time is confined by a resource factor~\cite{borrelli2017predictive}, in practice, which can model the battery and bandwidth of the network. This resource defined by $r\in\mathbb{R}^{n_r}$ is recharged at a constant rate $\rho$ until saturation, i.e., 
\[
\forall t\in [t_k,t_{k+1}),\;\;\dot{r}(t) = h(\overline{r}-r(t))\rho,
\]
where $\overline{r}$ is a saturation value and $h(\cdot)$ is the heaviside function with $h(s)=1$ if $s>0$ and $0$ elsewhere. When the agent is triggered to update the control input, the resource is discharged by an amount $\eta(\Delta_k)$ to pay the update cost. Thus, the resource at triggering time instants $\{t_k\}_{k=0}^{N-1}$ is
\begin{align}\label{eqn:resource_dyn}
r(t)=\begin{cases} r_0&\;\;t=t_0\\[0.16cm]
\lim\limits_{t\rightarrow t_k^-} r(t)-\eta(\Delta_k)&\;\; t\in\{t_k\}_{k=1}^{N-1}
\end{cases}
\end{align}
with an initially available resource  $r_0$ at $t_0$. Here, $t\to t_k^-$ represents the left limits, i.e., $t\to t_k$ and $t<t_k$. Moreover, the resource $r$ is further lower bounded by $r\in[\underline{r},\overline{r}]$. For the sake of compactness, we use the notation $v\in\mathbb{R}^{Nn_u}:=[v_0^\top,v_1^\top\dots,v_{N-1}^\top]^\top $ to stack the control coefficients, and
define the triggering time interval $\Delta_k:=t_{k+1}-t_k$ and use the notation $\Delta = [\Delta_0,...,\Delta_{N-1}]^\top$. And we denote $g(r(t_k),\Delta_k):=\min\left\{r(t_k)+\rho \Delta_k-\eta(\Delta_k),\overline{r}\right\}$. 

Accordingly, the resource-aware model predictive control (MPC) problem~\cite{wildhagen2020resource} can be summarized as 
\begin{subequations}\label{eqn:deter_resource_mpc}
\begin{align}
    \min_{x(\cdot),v,\Delta}\;\;& M(x(t_N))+\sum\limits_{k=0}^{N-1}\int_{t_{k}}^{t_{k+1}} l(x(\tau),v_k)d\tau\label{eq::obj}\\
    \text{s.t.}\quad\; &x(t_0)=x_0,\;r(t_0) = r_0,\label{eq::initial}\\
    &\forall\,t\in[t_0,t_N],\; \frac{dx(t)}{dt} = Ax(t)+Bu(t),\\
    &\forall\,t\in[t_0,t_N],\; x(t)\in\mathcal{X},\;u(t)\in\mathcal{U},\label{eq::stateCons}\\
    &\forall\,k\in\{0,1,...,N-1\}\notag\\
    &r(t_{k+1}) =g(r(t_k),\Delta_k),\label{eqn:discrete_resource_1}\\
    &r(t_{k+1})\in[\underline{r},\overline{r}],\label{eqn:resouce_cons}\\
    &\Delta_k\in[\underline{\Delta},\;\overline{\Delta}],\label{eqn:delta_cons}
\end{align}
\end{subequations}
where $l(\cdot,\cdot)$ and $M(\cdot)$ in~\eqref{eq::obj} are the stage and terminal costs respectively, the saturated resource dynamics~\eqref{eqn:discrete_resource_1} is a simplified yet equivalent formulation of the resource dynamics~\eqref{eqn:resource_dyn}~\cite{wildhagen2020resource}. The constraints of the triggering time interval in~\eqref{eqn:delta_cons} protect the system from becoming Zeno/frozen, meaning that the triggering time $\Delta$ is zero/infinite. The initial state and resource are given by~\eqref{eq::initial}.

In the receding horizon scheme, a resource-aware self-triggered agent can update its control input when its resource is sufficiently high to stay above the lower bound $\underline{r}$. Otherwise, it must wait until enough resource is available. Once the controller is triggered at the current time instance, the resource-aware self-triggered controller solves~\eqref{eqn:deter_resource_mpc} to plan the next triggering time and the associated control input.

\subsection{Differential Inequality}
Let us consider the uncertain continuous time autonomous dynamics $\dot{x}(t) = f(x(t),w(t))$ perturbed by $w\in\mathcal{W}$ for the compact set $\mathcal W\subset \mathbb R^{n_w}$. For a given set of initial states $X_0$ at $t_1$, we denote the reachable set at time $t_2>t_1$ as
\begin{align*}
    X(t_2) := \left\{\xi\in\R^{n_x} \middle\vert \begin{aligned}
         &\exists\;w(t)\in\mathcal{W},\;\forall\; t\in[t_1,t_2],  \\
         &\dot{x}(t) = f(x(t),w(t)),\\
         &x(t_1)\in X_0,\;x(t_2)=\xi.
    \end{aligned}\right\}.
\end{align*}
Moreover, we define the set-valued mapping
\begin{align*}
    \Gamma_f(c,X):=\left\{f(x,w)\middle\vert 
    \begin{aligned}
    c^\top\xi = V[X](c),\,x\in X,\, w\in\mathcal{W}
    \end{aligned}\right\}.
\end{align*}
The convex enclosure of the reachable set can be characterized by the following theorem.
\begin{theorem}\cite[Theorem 3]{villanueva2015unified}
\label{thm:diff_neq}
Let $Y:[t_1,t_2]\rightarrow \mathbb{K}^{n_x}_\mathrm{C}$ be a set-valued function such that
\begin{enumerate}
    \item \textit{the function $V[Y(\cdot)](c)$ is Lipschitz continuous on $[t_1,t_2]$ for all $c\in\R^{n_x}$and}
    \item \textit{the set-valued function $Y$ satisfies the differential inequality
    \begin{align*}
        \mathrm{a.e.}\;\; t\in[t_1,t_2],\; \dot{V}[Y(t)](c)\geq V[\Gamma_f(c,X)](c)
    \end{align*}
    with $V[Y(t_1)](c)\geq V[X_1](c)$ for all $c\in\R^{n_x}$.}
\end{enumerate}
Then, $Y$ is an enclosure of the reachable tube of $X(t)$ for all $t\in[t_1,t_2]$, \textit{i.e.}, $X(t)\subset Y(t),\;\forall\;t\in[t_1,t_2]$.
\end{theorem}

\subsection{Ellipsoidal Calculus}
This section recaps some useful results from ellipsoidal calculus~\cite{kurzhanskiui1997ellipsoidal}. The support function of an ellipsoid $\mathcal{E}(q,Q)$ is given by
\begin{align*}
    V[\mathcal{E}(q,Q)](c) = c^\top q+ \sqrt{c^\top Q c}\;.
\end{align*}
This value is obtained at the boundary of the ellipsoid as 
\begin{align*}
   Z[\mathcal{E}(q,Q)](c):= \text{arg}\max_z\{c^\top x|x\in\mathcal{E}(q,Q)\} = \frac{Qc}{\sqrt{c^\top Q c}}.
\end{align*}

The Minkowski sum of two ellipsoids is not necessarily an ellipsoid, and it can be outer approximated by $\forall\;\lambda\in(0,1)$,
\begin{align}
\label{eqn:sum_ellipse}
    \mathcal{E}(q_1,Q_2)\oplus\mathcal{E}(q_2,Q_2)\subset \mathcal{E}(q_1+q_2,\frac{Q_1}{\lambda}+\frac{Q_2}{1-\lambda}).
\end{align}

\section{Main Results}
\label{sect:main}
In this paper, we consider the following uncertain linear time-invariant dynamics
\begin{align}\label{eqn:dyn_uncertain}
    \frac{dx(t)}{dt} = Ax(t)+ B_uu(t)+B_w w(t)\;,
\end{align}
with matricies $B_u\in\mathbb{R}^{n_x\times n_u}$, $B_w\in\mathbb{R}^{n_x\times n_w}$, and uncertainty $w(t)\in\mathcal{E}(0,Q_w(t))$ for $Q_w(\cdot)\in\mathbb{S}_{++}^{n_w}$. In the following, we will derive the \textbf{continuous-time} dynamics of the ellipsoidal outer approximation of the reachable set $X(\cdot)$ driven by a \textbf{discrete-time} feedback control law. Notice that while the ellipsoidal outer approximation of a reachable set under \textbf{continuous-time} feedback control has been widely explored~\cite{Houska2019,kurzhanski2002reachability,brockman1998quadratic,schweppe1973uncertain}. However, because the triggering time is a decision variable in a self-triggered scheme, we observe that a direct application of most previous works is numerically unstable when used in an optimization algorithm. This motivates us to adopt differential inequality in this work. 

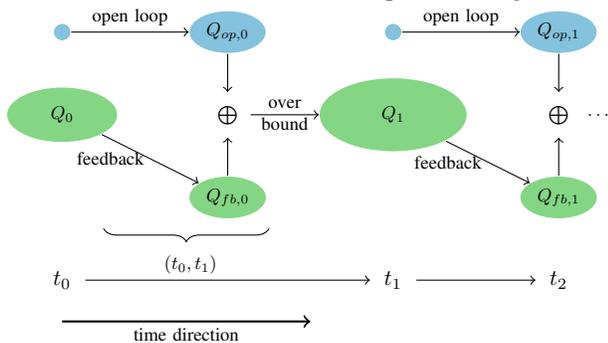
\begin{figure}[htbp!]
    \centering
    \begin{tikzpicture}[scale=1.1,every node/.style={scale=0.7}]
    \node[ellipse,draw = white, minimum height = 30pt, minimum width = 60pt,fill=MyGreen!80!white](Q0) at (0,0) {$Q_0$};
    \node[ellipse,draw = white, minimum height = 0.2pt, minimum width = .4pt,fill=MyBlue!60!white](Qop00) at (0,1) {};
    \node[ellipse,draw = white, minimum height = 15pt, minimum width = 15pt,fill=MyGreen!80!white](Qfb0) at (2,-1) {$Q_{fb,0}$};
    \node[ellipse,draw = white, minimum height = 0pt, minimum width = 15pt,fill=MyBlue!60!white](Qop0) at (2,1) {$Q_{op,0}$};
    \node[ellipse,draw = white, minimum height = 0.2pt, minimum width = .4pt,fill=MyBlue!60!white](Qop01) at (4,1) {};
    \node[circle,draw=white,fill=white,style={scale=1}] (sum0) at (2,0) {$\bigoplus$};
    \node[ellipse,draw = white, minimum height = 40pt, minimum width = 80pt,fill=MyGreen!80!white](Q1) at (4,0) {$Q_1$};
    \node[ellipse,draw = white, minimum height = 20pt, minimum width = 40pt,fill=MyGreen!80!white](Qfb1) at (6,-1) {$Q_{fb,1}$};
    \node[ellipse,draw = white, minimum height = 0pt, minimum width = 15pt,fill=MyBlue!60!white](Qop1) at (6,1) {$Q_{op,1}$};
    \node[circle,draw=white,fill=white,style={scale=1}] (sum1) at (6,0) {$\bigoplus$};
    \node[circle,draw=white,fill=white,style={scale=1}] (etc) at (6.5,0) {$\dots$};
    
    \node[circle,draw=white,fill=white,style={scale=1.2}] (t0) at (0,-2) {$t_0$};
    \node[circle,draw=white,fill=white,style={scale=1.2}] (t1) at (4,-2) {$t_1$};
    \node[circle,draw=white,fill=white,style={scale=1.2}] (t2) at (6,-2) {$t_2$};
    
    \draw [decorate,decoration={brace,amplitude=5pt,mirror,raise=4ex}]
  (0.5,-0.8) -- (2.5,-0.8) node[midway,yshift=-3em]{};
  \draw[thick,->] (0,-2.5) --node [midway, below,text centered] {time direction} (3,-2.5);
    
    \draw[thin,->] (Q0) --node [midway, left,text centered] {feedback} (Qfb0);
    \draw[thin,->] (Qfb0) -- (sum0);
    \draw[thin,->] (Qop0) -- (sum0);
    \draw[thin,->] (Qop00) --node [midway,above,text centered] {open loop} (Qop0);
    \draw[thin,->] (sum0) -- node [midway,text centered,text width=1cm] {over\\ bound}(Q1);
    \draw[thin,->] (Q1) --node [midway, left,text centered] {feedback} (Qfb1);
    \draw[thin,->] (Qop01) -- node [midway,above, text centered] {open loop}(Qop1);
    \draw[thin,->] (Qfb1) -- (sum1);
    \draw[thin,->] (Qop1) -- (sum1);
    \draw[thin, ->] (t0) --node [midway,above,text centered] {$(t_0,t_1)\qquad\qquad$} (t1);
    \draw[thin, ->] (t1) -- (t2);
    \end{tikzpicture}
    \caption{\label{fig:ellip_prop} Chain of ellipsoidal reachable set propagation (The tiny blue dots about $Q_0,\;Q_1$ is exaggerated to better depict that the accumulation of $Q_{op,0},\;Q_{op,1 }$ starts from zeros (See boundary condition in Lemma~\ref{lem:ellip_dyn}))}
\end{figure}

In self-triggered schemes, the control input is only allowed to change when the system is triggered. In particular, if the system is triggered at $t_k$ with its state contained in the reach set $X(t_k)$, which in turn is bounded within $\mathcal{E}(q_k,Q_k)$, we propose to update its input via a nominal term $v_k$ and a feedback term $K$ as $\forall\, t\in(t_k,t_{k+1}]$,
\begin{align}\label{eqn:input}
    u_k(t,x) = v_k+Kx,\;\;x\in\mathcal{E}(q_k,Q_k).
\end{align}
This control input must then remain constant until its next trigger at $t_{k+1}$. Before delving into the details of the proposed controller, we summarize the mechanism to first give an intuitive general viewpoint. Given any sequence of $\{v_i\}_{i=0}^{N-1}$, $\{t_i\}_{i=0}^{N}$ and a feedback control law $K$,  we can define the chain of reachable sets depicted in Figure~\ref{fig:ellip_prop}. 

Without loss of generality, we consider the triggering time instances $\{t_k\}_{k=0}^2$ shown in this figure, and the following two ingredients govern the evolution of the reachable set:
\begin{itemize}
    \item \textbf{The reachable set dynamics between two consecutive triggers:} Consider the time interval $(t_0,t_1)$, the reachable set dynamics are decomposed into a closed-loop (\textit{i.e.} feedback) component (the path between the green ellipsoids $Q_0$ and $Q_{fb,0}$ in Figure~\ref{fig:ellip_prop}) and an open-loop component (the path between the blue dot and $Q_{op,0}$ in Figure~\ref{fig:ellip_prop}), each of which evolves independently due to the self-triggered mechanism. 
    These dynamics will be summarized in Lemma~\ref{lem:ellip_dyn}.
    \item \textbf{The reachable set outer-approximation at triggering time instances:} Consider the trigger at $t_1$, the reachable set developed over the interval $(t_0,t_1)$ is outer approximated by an ellipsoid $\mathcal{E}(q_1,Q_1)$ (Big green ellipsoid $Q_1$ in Figure~\ref{fig:ellip_prop}).
\end{itemize}

Based on the discussion above, characterization of the reachable set propagation between two consecutive triggers is vital to the MPC design. The propagation of the reachable set outer approximation between two consecutive triggers $(t_k,t_{k+1}]$ is stated in the following.
\begin{lemma}\label{lem:ellip_dyn}
Let $X(t_k)\subseteq\mathcal{E}(q_k,Q_k)$ and dynamics~\eqref{eqn:dyn_uncertain} be driven by control law~\eqref{eqn:input}. The reachable set $X(t)$ for all $t\in[t_k,t_{k+1}]$ is outer bounded by
\begin{align}\label{eqn:outer_decomp}
    X(t)\subseteq \mathcal{E}(q_{k}(t),Q_{fb,k}(t))\oplus \mathcal{E}(0,Q_{op,k}(t,\lambda_k(t))),
\end{align}
where $\lambda_k(\cdot):\mathbb{R}\rightarrow\mathbb{R}^+$ is any positive real-valued function on $[t_k,t_{k+1}]$ and the shape of the outer approximation is characterized by 
\begin{subequations}\label{eqn:outer_bound_dyn}
\begin{align}
    \frac{dq_{k}(t)}{dt} =&\; Aq_{k}(t)+Bv_k\label{eqn:outer_fb_mean}\\
    \frac{dQ_{fb,k}(t)}{dt} =&\; AQ_{fb,k}(t) +Q_{fb,k}(t)A^\top\nonumber\\
    +B_u&KQ_{cr,k}(t)^\top+Q_{cr,k}(t)(B_uK)^\top\label{eqn:outer_fb_psd1}\\
    \frac{dQ_{cr,k}(t)}{dt} = &\; AQ_{cr,k}(t)+BK Q_{k}\label{eqn:outer_fb_psd2}\\
    \frac{dQ_{op,k}(t,\lambda_k(t))}{dt} = &\; AQ_{op,k}(t,\lambda_k(t))+Q_{op,k}(t,\lambda_k(t))A^\top\nonumber\\
    +\lambda_k(t)&Q_{op,k}(t,\lambda_k(t))+\frac{B_wQ_w(t)B_w^\top}{\lambda_k(t)}\label{eqn:outer_op}
\end{align}
\end{subequations}
with $Q_{fb,k}(t_k)=Q_{cr,k}(t_k)=Q_k$, $Q_{op,k}(t_k,\lambda_k(t_k))=\mathrm{\bf 0}$.
\end{lemma}
{\small
\textbf{Proof.}
In our proof, we first derive the decomposition in~\eqref{eqn:outer_decomp} and then, work out the dynamics~\eqref{eqn:outer_bound_dyn} for $k$-th interval $[t_k,t_{k+1}]$.

\textbf{Reachable set decomposition:}
Note that the reachable set driven by control law~\eqref{eqn:input} is 
\begin{align}\notag
    &X(t):=\\\notag
    &\;\; \left\{\xi\in\R^{n_x} \middle\vert 
    \begin{aligned}
    &\exists\,w(\tau)\in\mathcal{E}(0,Q_w(\tau)),\;\forall\, \tau\in[t_k,t],\\
    &\dot{x}(\tau) = Ax(\tau)+B_u(v_k+K\Tilde{x})+B_ww(\tau)\\
    &x(t_k) = \Tilde{x}\in\mathcal{E}(q_k,Q_k),\;x(t)=\xi
    \end{aligned}\right\}\\[0.16cm]\label{eq::Eq_a}
    &= \underbrace{\left\{\xi\in\R^{n_x} \middle\vert \begin{aligned}
    &\dot{x}(\tau) = Ax(\tau)+B_u(v_k+K\Tilde{x})\\
    &x(t_k) = \Tilde{x}\in\mathcal{E}(q_k,Q_k),\;x(t)=\xi\\
    \end{aligned}\right\}}_{X_{fb,k}(t)}\\\notag
    &\;\;\;\oplus \underbrace{\left\{\xi\in\R^{n_x} \middle\vert \begin{aligned}
    &\exists\;w(\tau)\in\mathcal{E}(0,Q_w(\tau)),\;\forall\; \tau\in[t_k,t],  \\
    &\dot{x}(\tau) = Ax(\tau)+B_ww(\tau),\\
    &x(t_k)=0\;,\;x(t)=\xi
    \end{aligned}\right\}}_{X_{op,k}(t)},
\end{align}
where Equality~\eqref{eq::Eq_a} follows the linearity of the dynamics~\eqref{eqn:dyn_uncertain}.

\textbf{Dynamics of feedback component:}
The set $X_{fb,k}(t)$ can be rewritten as
\begingroup\makeatletter\def\f@size{8.8}\check@mathfonts
\begin{align}\notag
    X_{fb,k}&(t):= \left\{\xi\in\R^{n_x} \middle\vert \begin{aligned}
        &\exists\; z\in \R^{n_x},\;z^\top z\leq 1,\;\forall\, \tau\in[t_k,t],\\
         &\dot{x}(\tau) = Ax(\tau)+B_u(v_k+K\Tilde{x})\\
         &x(t_k) = \Tilde{x}= q_k+Q_k^{\frac{1}{2}}z,\;x(t)=\xi\;\;\\
    \end{aligned}\right\}\\\label{eq::Eq_b}
     &= \left\{\begin{aligned}
     &\underbrace{e^{At}q_k+\int_0^te^{A(t-\tau)}B_uv_kd\tau}_{(a)}\\
     &+\underbrace{\int_0^te^{A(t-\tau)}B_ud\tau(KQ^{\frac{1}{2}}z)}_{(b)}\end{aligned}\middle\vert z^\top z\leq 1\right\},
\end{align}
\endgroup
where~\eqref{eq::Eq_b} utilizes the explicit solution of $\dot x(t) = Ax(t)+Bu(t)$. Notice that, by definition of an ellipsoid, $X_{fb,k}(t)$ is also an ellipsoid, whose center is given by the term $(a)$ and the shape is defined by the term $(b)$. 
For simplification, we denote it by $\mathcal{E}(q_{k}(t),Q_{fb,k}(t))$. In particular,
\begin{align*}
    &q_{k}(t) = e^{At}q_k+\int_0^te^{A(t-\tau)}B_ud\tau v_k\\
    \Longrightarrow&\;\frac{d q_{k}(t)}{dt}= A q_{k}(t)+B_u v_k\;,
\end{align*}
which is the dynamics~\eqref{eqn:outer_fb_mean}. Moreover, we denote $\Tilde{B}(t):=\int_0^te^{A(t-\tau)}d\tau B_u$ such that we have 
\begin{align*}
    &Q_{fb,k}(t) = \Tilde{B}(t)KQ_k\left(\Tilde{B}(t)K\right)^\top\\
    \Longrightarrow\;&\frac{dQ_{fb,k}(t)}{dt} = AQ_{fb,k}(t)+Q_{fb,k}(t)A^\top\\
    &\qquad\qquad+B_uKQ_k\left(\Tilde{B}(t)K\right)^\top + \underbrace{\Tilde{B}(t)KQ_k}_{Q_{cr,k}(t)}(B_uK)^\top\\
    &\frac{dQ_{cr,k}(t)}{dt} = AQ_{cr,k}(t)+B_uKQ_k\;,
\end{align*}
which recovers~\eqref{eqn:outer_fb_psd1} and~\eqref{eqn:outer_fb_psd2}.

\textbf{Dynamics of open-loop component:}
The remaining proof will construct an outer approximation of $X_{op,k}(t)$ with ellipsoid $\mathcal{E}(0,Q_{op,k}(t,\lambda_k(t)))$. Note that the autonomous dynamics considered in the reachable set $X_{op,k}(t)$ are $$\dot{x}(t)=f_w(x,w):=Ax(t)+B_w w(t).$$ 
In order to apply Theorem~\ref{thm:diff_neq}, we introduce support function
\begin{align*}
    &V[\Gamma_{f_w}(c,\mathcal{E}(0,Q_{op,k}(t,\lambda_k(t)))\,)](c) \\
    =\,& \max_{w\in\mathcal{E}(0,Q_w(t))}c^\top \left(A Z[\mathcal{E}(0,Q_{op,k}(t,\lambda_k(t)))](c))+Bw\right)\\
    =\,&\max_{w\in\mathcal{E}(0,Q_w(t))} c^\top\left(A\frac{Q_{op,k}(t,\lambda_k(t))c}{\sqrt{c^\top Q_{op,k}(t,\lambda_k(t))c}}+B_ww\right)\\
    =\,&c^\top A\frac{Q_{op,k}(t,\lambda_k(t))c}{\sqrt{c^\top Q_{op,k}(t,\lambda_k(t))c}} + \sqrt{c^\top B_wQ_w(t)B_w^\top c}\;.
\end{align*}
We apply Theorem~\ref{thm:diff_neq} to outer approximate the shape of the $\mathcal{E}(0,Q_{op,k}(t,\lambda_k(t)))$, which yields
\begin{equation*}
\begin{split}
    &\dot V[\mathcal{E}(0,Q_{op,k}(t,\lambda_k(t)))](c)\geq\\
    &\qquad\qquad\max_{w\in\mathcal{E}(0,Q_w)} c^\top\left(A\frac{Q_{op,k}(t,\lambda_k(t))c}{\sqrt{c^\top Q_{op,k}(t,\lambda_k)c}}+B_ww\right)\\
    \Longrightarrow&\frac{1}{2}c^\top \dot{Q}_{op,k}(t,\lambda_k(t))c\geq c^\top AQ_kc \\  
    &\qquad\qquad\;\;+\sqrt{c^\top B_wQ_w(t)B_w^\top c}{\sqrt{c^\top Q_{op,k}(t,\lambda_k(t))c}}.\\
\end{split}
\end{equation*}
By applying the tight arithmetic-geometric mean inequality~\cite{steele2004cauchy}, we reformulate the second inequality above as 
\begin{equation}
\label{eq::ineq_c}
\begin{aligned}
    &\frac{1}{2}c^\top \dot{Q}_{op,k}(t,\lambda_k(t))c\\
    \geq& c^\top A Q_k c + 
    \inf\limits_{\lambda>0}\;\;\frac{1}{2\lambda}c^\top B_wQ_wB_w^\top c + \frac{\lambda}{2}c^\top Q_{op,k}c\,.
\end{aligned}
\end{equation}

According to Theorem~\ref{thm:diff_neq}, we can construct a ellipsoidal outer approximation of $X_{op,k}(t)$ by enforcing the following inequality
\begin{equation*}
    \begin{split}
    &c^\top \dot{Q}_{op,k}(t,\lambda_k(t))c \\
    =& \underbrace{\frac{1}{2}\left(c^\top \dot{Q}_{op,k}(t,\lambda_k(t))c+(c^\top \dot{Q}_{op,k}(t,\lambda_k(t))c)^\top\right)}_{(d)}\\
    \geq& \inf\limits_{\lambda>0} c^\top\left( \begin{aligned}
    &AQ_{op,k}(t,\lambda_k(t))+ Q_{op,k}(t,\lambda_k(t))A^\top\\
    &+\frac{B_wQ_w(t)B_w^\top}{\lambda_k(t)} +\lambda_k(t) Q_{op,k}(t,\lambda_k(t))\end{aligned}\right)c\;,
\end{split}
\end{equation*}
where the decomposition in $(d)$ is used to build a symmetric form of $Q_{op,k}(t)$ from the asymmetric form that appeared in~\eqref{eq::ineq_c}. The final step is to get rid of the inequality and the infimum operator. Note that if there exists $\lambda_k(\cdot):\R\rightarrow\R^+$ such that the following dynamics is satisfied, then the inequality~\eqref{eq::ineq_c} is satisfied. 
\begin{align*}
    \frac{dQ_{op,k}(t,\lambda_k(t))}{dt} =&  AQ_{op,k}(t,\lambda_k(t))+Q_{op,k}(t,\lambda_k(t))A^\top\\
    &+\frac{B_wQ_w(t)B_w^\top}{\lambda_k(t)}+\lambda_k(t) Q_{op,k}(t,\lambda_k(t)).
\end{align*}
This recovers the dynamics~\eqref{eqn:outer_op}, which results in the inclusion $X_{op,k}(t)\subset\mathcal{E}(0,Q_{op,k}(t))$ by construction, we thus have 
\[
\begin{aligned}
    X(t)&\subseteq \mathcal{E}(q_{fb,k}(t),Q_{fb,k}(t))\oplus X_{op,k}(t)\\
    &\subset \mathcal{E}(q_{fb,k}(t),Q_{fb,k}(t))\oplus\mathcal{E}(0,Q_{op,k}(t,\lambda_k(t)))
\end{aligned}
\]
for any $t\in[t_k,t_{k+1}]$, which concludes the decomposition~\eqref{eqn:outer_decomp}.
}\hfill$\blacksquare$
\begin{remark}
Lemma~\ref{lem:ellip_dyn} indicates that the evolution of the uncertainty between two consecutive triggers $[t_k,t_{k+1}]$ can be decomposed into two dynamic parts. Both are independently driven by the zero-order-hold feedback generated at $t_k$ and by the open-loop accumulation of the disturbance $w(t)$, respectively. The former corresponds to $\mathcal{E}(q_k(t),Q_{fb,k}(t))$ and the latter is an outer approximation given by $\mathcal{E}(0,Q_{op,k}(t,\lambda_k(t)))$. \change{Moreover, as the trigger occurs at $t_k$, $Q_{cr,k}(t)$ can be perceived as the correlation between the uncertainty at $t$ and $t_k$, see~\eqref{eqn:outer_fb_psd2}. This, in terms, reflects the self-triggered property.}
\end{remark}

\subsection{Robust Resource-Aware MPC}
This section summarizes a robust MPC controller that incorporates the dynamics derived in the last section into the self-triggered MPC scheme. In particular, the controller optimizes the nominal performance while ensuring a robust input/state constraint satisfaction. In general, the nominal inputs $\{v_k\}_{k=0}^{N-1}$, the feedback control $K$, and the triggering time instances $\{t_k\}_{k=1}^N$ are determined by solving the following problem:
\[
\underset{\substack{K,v,\Delta,q_{fb},\lambda(\cdot)\\Q_{fb},Q_{cr},Q_{op},\kappa}}{\text{minimize}}\;\;M(q(t_N))+\sum_{k=0}^{N-1}\int_{t_{k}}^{t_{k+1}} l(q_{fb,k}(\tau),v_k)d\tau 
\]
subject to 
\begin{subequations}\label{eqn:rb_resource_mpc}
\small
\begin{align}
    &X(t_0)=\mathcal{E}(q_0,Q_0),\;r(t_0) = r_0,\label{eqn::rb_mpc_init}\\
    &\left\{\begin{aligned}
    &\forall\,t\;\in(t_0,t_N),\;\forall\;k\in\mathbb{Z}_0^{N-1},\\
    &X(t)\subseteq \mathcal{X}, v_k+KX(t_k)\subseteq \mathcal{U},\;r(t_{k+1})\in[\underline{r},\overline{r}]\\
    &r(t_{k+1}) = g(r(t_k),\Delta_k),\;\Delta_k\in[\underline{\Delta},\overline{\Delta}],
    \end{aligned}\right.\\
    &\left\{\begin{aligned}
    &\forall\; t\in[t_k,t_{k+1}],\;\forall\;k\in\mathbb{Z}_0^{N-1},\\
    &X(t)\subseteq\mathcal{E}(q_{k}(t),Q_{fb,k}(t))\oplus \mathcal{E}(0,Q_{op,k}(t,\lambda_k(t))),\\
    &\frac{dq_{fb,k}(t)}{dt} =\; Aq_{fb,k}(t)+Bv_k\;,q_{fb,k}(t_k)=q_k,\\
    &\frac{dQ_{fb,k}(t)}{dt} =\; AQ_{fb,k}(t) +Q_{fb,k}(t)A^\top,\\
    &\qquad +B_uKQ_{cr,k}(t)^\top+Q_{cr,k}(t)(B_uK)^\top,\\
    &\frac{dQ_{cr,k}(t)}{dt} = \; AQ_{cr,k}(t)+BK Q_{i},\\
    &\frac{dQ_{op,k}(t,\lambda_k(t))}{dt} = \; AQ_{op,k}(t,\lambda_k(t)),\\
    &\qquad +Q_{op,k}(t,\lambda_k(t))A^\top+\lambda_k(t) Q_{op,k}(t,\lambda_k(t)),\\
    &\qquad+\frac{B_wQ_w(t)B_w^\top}{\lambda_k(t)},\;\;Q_{op,k}(t_k,\lambda_k(t_k))=\mathbf{0},\\
    &Q_{fb,k}(t_k)=Q_{cr,k}(t_k)=Q_k,
    \end{aligned}\right.\label{eqn:rb_dyn}\\
    &\left\{
    \begin{aligned}
    &\forall\; t_k\;\text{ with }\; k\in\mathbb{Z}_1^{N-1}:\\
    &Q_k = \frac{Q_{fb,k-1}(t_k)}{\kappa_k}+\frac{Q_{op,k-1}}{1-\kappa_k}\;,\;\kappa_k\in(0,1).
    \end{aligned}
    \right.\label{eqn:continuity}
\end{align}
\end{subequations}
The initial conditions are enforced by~\eqref{eqn::rb_mpc_init} with potentially uncertain measurements. The dynamics summarized in Lemma~\ref{lem:ellip_dyn} is enforced in constraints~\eqref{eqn:rb_dyn}. \eqref{eqn:continuity} models how the reachable sets from adjacent intervals connect to each other (see the over bound arrow in the middle of Figure~\ref{fig:ellip_prop}). In particular, the reachable set in time interval $[t_{k-1},t_{k}]$ links to the reachable set in $[t_{k},t_{k+1}]$ at $t_{k}$, where the ellipsoid $Q_k$ is used to generate the outer approximation of 
$$\mathcal{E}(q_{k}(t),Q_{fb,k}(t))\oplus \mathcal{E}(0,Q_{op,k}(t,\lambda_k(t))).$$ 
Finally, we summarize a few important notes to enable an efficient implementation of the proposed MPC controller~\eqref{eqn:rb_resource_mpc}. 
\begin{itemize}[leftmargin=*]
    \item To solve the problem within a direct optimal control scheme, the integration of the ordinary differential equations can be achieved by numerical integration methods such as the Runge-Kutta method or the collocation method~\cite{levine2018handbook}. In this case, the collocation method is preferable because the integration is linear with respect to the triggering time difference $\{\Delta_k\}_{k=0}^{N-1}$, while other numerical integration methods depend on high order terms of $\{\Delta_k\}_{k=0}^{N-1}$, which results in low numerical stability.
    \item When the feasible sets $\mathcal{X},\;\mathcal{U}$ are ellipsoidal, equation~\eqref{eqn:sum_ellipse} can be used to determine the satisfaction of both constraints. If these sets are polytopic, then calculus of support functions can be applied. Each linear constraint can be re-written as
\end{itemize}
\[
\begin{aligned}
    &\forall\,x\in\mathcal{E}(q_{k}(t),Q_{fb,k}(t))\oplus \mathcal{E}(0,Q_{op,k}(t,\lambda_k(t))),\;c^\top x\leq C,\\
    &\Longrightarrow c^\top q_k+V[\mathcal{E}(0,Q_{fb,k}(t))](c)\\
    &\qquad\qquad\qquad\quad\qquad+V[\mathcal{E}(0,Q_{op,k}(t,\lambda_k(t)))](c)\leq C.
\end{aligned}
\]
\begin{remark}
\change{
If the sampling time is fixed, the proposed robust controller will coincide with the discrete-time ellipsoidal robust MPC method by discretizing the continuous time dynamics~\eqref{eqn:dyn_uncertain} explicitly. 
To see this, recall that the first component in~\eqref{eqn:outer_decomp}, $\mathcal{E}(q_k(t),Q_{fb,k}(t))$ is derived from the discrete-time explicit solution~\eqref{eq::Eq_b}. Meanwhile, the second component in~\eqref{eqn:outer_decomp}, $\mathcal{E}(\textbf{0},Q_{op,k}(t,\lambda_k(t)))$ coincides with the ellipsoidal additive process noise generated by explicit discretization~\cite[Theorem 5.1 and Remark 5.2]{houska2011robust}. Hence, the proposed robust controller will not introduce extra conservativeness in comparison with the state-of-art discrete-time ellipsoidal robust MPC.}
\end{remark}

\section{Numerical Results}\label{sect:num}
The proposed algorithm has been tested on a double integrator with state $x(t)=(x_1(t),x_2(t))$, whose dynamics are
\[
\frac{dx(t)}{dt} =\begin{bmatrix}0 &1\\ 0&0\end{bmatrix}x(t)+\begin{bmatrix}0\\1\end{bmatrix}u(t)+\begin{bmatrix}0\\1\end{bmatrix}w(t).
\]
The controller is designed to track a reference signal oscillating between $1$ and $-0.2$ by choosing the stage cost to be $l(x(t),u(t))= 10(x_1(t)-x^\mathrm{ref}(t))^2$, where $x^\mathrm{ref}$ is the tracking reference. The recharging rate is $0.8$ with a trigger cost of $0.4$. To show the effectiveness of the proposed algorithm, we consider two different cases. In the first case, the disturbance $w(t)$ is bounded within $[-0.04,0.04]$, and in the second case, $w\in[-0.2,0.2]$. In both cases, we consider an input constraint of $[-5,5]$ and an output constraint of $\begin{bmatrix}-1\\-10\end{bmatrix}\leq x\leq \begin{bmatrix}1\\10\end{bmatrix}$ with a prediction horizon $N=8$. The triggering time is bounded within $[0.1,1.5]$ with a resource constraint $r\in[0,1]$. \change{In both experiments, we also compare the proposed scheme to a closed-loop robust MPC with fixed sampling time, and we set the sampling frequence as high as possible regarding the resource dynamics (\textit{i.e. $\Delta = \mu/\rho=0.5s$}).}

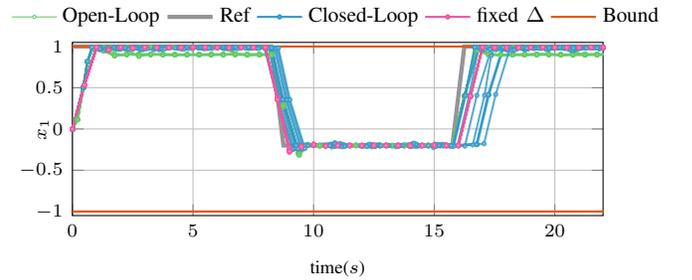
\begin{figure}[htbp!]
    \centering
    \begin{tikzpicture}
    \begin{axis}[xmin=0, xmax=22,
    ymin=-1.05, ymax= 1.05,
    enlargelimits=false,
    clip=true,
    grid=major,
    mark size=0.5pt,
    width=1\linewidth,
    height=0.45\linewidth,ylabel = $x_1$,xlabel= time($s$),
    legend style={
    	font=\footnotesize,
    	draw=none,
		at={(0.5,1.03)},
        anchor=south
    },
    legend columns=5,
    label style={font=\scriptsize},
    ylabel style={at={(axis description cs:0.12,0.5)}},
    xlabel style={at={(axis description cs:0.5,0.02)}},
    ticklabel style = {font=\scriptsize}]
    
    \pgfplotstableread{data/open_loop/1.dat}{\dat}
    \addplot+ [thin, mark=*, mark options={fill=white, scale=1.2},MyGreen] table [x={t}, y={y}] {\dat};
    \addlegendentry{Open-Loop}
    \addplot+ [ultra thick,Gray,mark=none] table [x={t}, y={ref}] {\dat};
    \addlegendentry{Ref}
    \addlegendimage{line legend,thick, mark=*, mark options={fill=white, scale=1.2}, MyBlue}
    \addlegendentry{Closed-Loop} 
    \addlegendimage{line legend,thick, mark=*, mark options={fill=white, scale=1.2}, VioletRed}
    \addlegendentry{fixed $\Delta$}
    \addplot+ [thick, MyRed, mark = none] table [x={t}, y={max}] {\dat};
    \addplot+ [thick, MyRed, mark = none] table [x={t}, y={min}] {\dat};
    \addlegendentry{Bound}
    
    \pgfplotstableread{data/open_loop/2.dat}{\dat}
    \addplot [thick, mark=*, mark options={fill=white, scale=1.2}, MyGreen] table [x={t}, y={y}] {\dat};
    
    \pgfplotstableread{data/open_loop/3.dat}{\dat}
    \addplot [thick, mark=*, mark options={fill=white, scale=1.2}, MyGreen] table [x={t}, y={y}] {\dat};
    
    \pgfplotstableread{data/open_loop/4.dat}{\dat}
    \addplot [thick, mark=*, mark options={fill=white, scale=1.2}, MyGreen] table [x={t}, y={y}] {\dat};
    
    \pgfplotstableread{data/open_loop/5.dat}{\dat}
    \addplot [thick, mark=*, mark options={fill=white, scale=1.2},MyGreen] table [x={t}, y={y}] {\dat};
    
    \pgfplotstableread{data/open_loop/6.dat}{\dat}
    \addplot [thick, mark=*, mark options={fill=white, scale=1.2},MyGreen] table [x={t}, y={y}] {\dat};
    
    \pgfplotstableread{data/open_loop/7.dat}{\dat}
    \addplot [thick, mark=*, mark options={fill=white, scale=1.2},MyGreen] table [x={t}, y={y}] {\dat};
    
    \pgfplotstableread{data/open_loop/8.dat}{\dat}
    \addplot [thick, mark=*, mark options={fill=white, scale=1.2},MyGreen] table [x={t}, y={y}] {\dat};
    
    \pgfplotstableread{data/open_loop/9.dat}{\dat}
    \addplot [thick, mark=*, mark options={fill=white, scale=1.2},MyGreen] table [x={t}, y={y}] {\dat};
    
    \pgfplotstableread{data/open_loop/10.dat}{\dat}
    \addplot [thick, mark=*, mark options={fill=white, scale=1.2},MyGreen] table [x={t}, y={y}] {\dat};
    
    \pgfplotstableread{data/closed_loop/004/1.dat}{\dat}
    \addplot [thick,opacity = 0.8, mark=*, mark options={fill=white, scale=1.2}, MyBlue] table [x={t}, y={y}] {\dat};
        \pgfplotstableread{data/closed_loop/004/2.dat}{\dat}
    \addplot [thick,opacity = 0.8,mark=*, mark options={fill=white, scale=1.2}, MyBlue] table [x={t}, y={y}] {\dat};
        \pgfplotstableread{data/closed_loop/004/3.dat}{\dat}
    \addplot [thick,opacity = 0.8, mark=*, mark options={fill=white, scale=1.2}, MyBlue] table [x={t}, y={y}] {\dat};
        \pgfplotstableread{data/closed_loop/004/4.dat}{\dat}
    \addplot [thick,opacity = 0.8, mark=*, mark options={fill=white, scale=1.2}, MyBlue] table [x={t}, y={y}] {\dat};
        \pgfplotstableread{data/closed_loop/004/5.dat}{\dat}
    \addplot [thick,opacity = 0.8,mark=*, mark options={fill=white, scale=1.2}, MyBlue] table [x={t}, y={y}] {\dat};
        \pgfplotstableread{data/closed_loop/004/6.dat}{\dat}
    \addplot [thick,opacity = 0.8, mark=*, mark options={fill=white, scale=1.2}, MyBlue] table [x={t}, y={y}] {\dat};
        \pgfplotstableread{data/closed_loop/004/7.dat}{\dat}
    \addplot [thick,opacity = 0.8, mark=*, mark options={fill=white, scale=1.2}, MyBlue] table [x={t}, y={y}] {\dat};
        \pgfplotstableread{data/closed_loop/004/8.dat}{\dat}
    \addplot [thick,opacity = 0.8, mark=*, mark options={fill=white, scale=1.2}, MyBlue] table [x={t}, y={y}] {\dat};
        \pgfplotstableread{data/closed_loop/004/9.dat}{\dat}
    \addplot [thick,opacity = 0.8, mark=*, mark options={fill=white, scale=1.2}, MyBlue] table [x={t}, y={y}] {\dat};
        \pgfplotstableread{data/closed_loop/004/10.dat}{\dat}
    \addplot [thick,opacity = 0.8, mark=*, mark options={fill=white, scale=1.2}, MyBlue] table [x={t}, y={y}] {\dat};
    
    \pgfplotstableread{data/fixed_dt/004/1.dat}{\dat}
    \addplot [thick,opacity = 0.8, mark=*, mark options={fill=white, scale=1.2}, VioletRed] table [x={t}, y={y}] {\dat};
        \pgfplotstableread{data/fixed_dt/004/2.dat}{\dat}
    \addplot [thick,opacity = 0.8,mark=*, mark options={fill=white, scale=1.2}, VioletRed] table [x={t}, y={y}] {\dat};
        \pgfplotstableread{data/fixed_dt/004/3.dat}{\dat}
    \addplot [thick,opacity = 0.8, mark=*, mark options={fill=white, scale=1.2}, VioletRed] table [x={t}, y={y}] {\dat};
        \pgfplotstableread{data/fixed_dt/004/4.dat}{\dat}
    \addplot [thick,opacity = 0.8, mark=*, mark options={fill=white, scale=1.2}, VioletRed] table [x={t}, y={y}] {\dat};
        \pgfplotstableread{data/fixed_dt/004/5.dat}{\dat}
    \addplot [thick,opacity = 0.8,mark=*, mark options={fill=white, scale=1.2}, VioletRed] table [x={t}, y={y}] {\dat};
        \pgfplotstableread{data/fixed_dt/004/6.dat}{\dat}
    \addplot [thick,opacity = 0.8, mark=*, mark options={fill=white, scale=1.2}, VioletRed] table [x={t}, y={y}] {\dat};
        \pgfplotstableread{data/fixed_dt/004/7.dat}{\dat}
    \addplot [thick,opacity = 0.8, mark=*, mark options={fill=white, scale=1.2}, VioletRed] table [x={t}, y={y}] {\dat};
        \pgfplotstableread{data/fixed_dt/004/8.dat}{\dat}
    \addplot [thick,opacity = 0.8, mark=*, mark options={fill=white, scale=1.2}, VioletRed] table [x={t}, y={y}] {\dat};
        \pgfplotstableread{data/fixed_dt/004/9.dat}{\dat}
    \addplot [thick,opacity = 0.8, mark=*, mark options={fill=white, scale=1.2}, VioletRed] table [x={t}, y={y}] {\dat};
        \pgfplotstableread{data/fixed_dt/004/10.dat}{\dat}
    \addplot [thick,opacity = 0.8, mark=*, mark options={fill=white, scale=1.2}, VioletRed] table [x={t}, y={y}] {\dat};
    
    \end{axis}
    \end{tikzpicture}
    \caption{\change{Comparison: open-loop vs closed-loop robust self triggered MPC}}
    \label{fig:compare_output}
\end{figure} 
In the first case, we further compare the closed-loop scheme with the open-loop scheme (\textit{i.e.}, $K=\mathbf{0}$). A relatively small process noise is considered ($w(t)\in[-0.04,0.04]$) in 20 runs of Monte-Carlo test, whose output responses are plotted in Figure~\ref{fig:compare_output}. These three controllers guarantee robust output constraint satisfaction. In comparison with the open-loop scheme, the proposed controller shows tacking performance when the reference signal and  the output upper bound overlap at $1$. More frequent trigger is also observed in the open-loop scheme (Figure~\ref{fig:compare_resource}), as resource consumption of the closed-loop controller is much lower, which accordingly implies that the sensors/CPUs in the closed-loop controller can stay longer in idle/deep-sleep mode to save more energy. This phenomenon is more significant when the reference signal is at $-0.2$, which is distant from the output constraints, the closed-loop controller quickly recharges its resources while the open-loop controller still actuates at the highest frequency. It is noteworthy to point out that longer idle mode does not necessarily lead to a better performance. For example, in the step change at around 11 seconds in Figure~\ref{fig:compare_output}, the closed-loop controller does not response to the reference change as it is still in idle mode. However, these problems can be fixed by forced \change{triggering} the agent when a reference change is detected. 
\begin{figure}[t]
    \centering
    \begin{tikzpicture}
    \begin{axis}[xmin=0, xmax=22,
    ymin= 0 , ymax= 1.05,
    enlargelimits=false,
    clip=true,
    grid=major,
    mark size=0.5pt,
    width=1\linewidth,
    height=0.39\linewidth,ylabel = $r$,xlabel= time($s$),
    legend style={
    	font=\footnotesize,
    	draw=none,
		at={(0.5,1.03)},
        anchor=south
    },
    legend columns=3,
    label style={font=\scriptsize},
    ylabel style={at={(axis description cs:0.06,0.5)}},
    xlabel style={at={(axis description cs:0.5,0.02)}},
    ticklabel style = {font=\scriptsize}]
    
    \pgfplotstableread{data/open_loop/1.dat}{\dat}
    \addplot [ybar,bar width=3pt, mark=none, MyGreen, solid] table [x={t}, y={r}] {\dat};
    \addlegendentry{Open-Loop}
    \addlegendimage{line legend,ultra thick, mark=none, MyBlue}
    \addlegendentry{Closed-Loop}  
    \addplot+ [thick, MyRed, mark = none] table [x={t}, y={rmax}] {\dat};
    \addplot+ [thick, MyRed, mark = none] table [x={t}, y={rmin}] {\dat};
    \addlegendentry{Resource Constraints}
    
    \pgfplotstableread{data/open_loop/2.dat}{\dat}
    \addplot [ybar,bar width=2pt, mark=none, MyGreen, solid] table [x={t}, y={r}] {\dat};
    
    \pgfplotstableread{data/open_loop/3.dat}{\dat}
    \addplot [thick, mark=*, mark options={fill=white, scale=1.2}, MyGreen] table [x={t}, y={r}] {\dat};
    
    \pgfplotstableread{data/open_loop/4.dat}{\dat}
    \addplot [ybar,bar width=2pt, mark=none, MyGreen, solid] table [x={t}, y={r}] {\dat};
    
    \pgfplotstableread{data/open_loop/5.dat}{\dat}
    \addplot [ybar,bar width=2pt, mark=none, MyGreen, solid] table [x={t}, y={r}] {\dat};
    
    \pgfplotstableread{data/open_loop/6.dat}{\dat}
    \addplot [ybar,bar width=2pt, mark=none, MyGreen, solid] table [x={t}, y={r}] {\dat};
    
    \pgfplotstableread{data/open_loop/7.dat}{\dat}
    \addplot [ybar,bar width=2pt, mark=none, MyGreen, solid] table [x={t}, y={r}] {\dat};
    
    \pgfplotstableread{data/open_loop/8.dat}{\dat}
    \addplot [ybar,bar width=2pt, mark=none, MyGreen, solid] table [x={t}, y={r}] {\dat};
    
    \pgfplotstableread{data/open_loop/9.dat}{\dat}
    \addplot [ybar,bar width=2pt, mark=none, MyGreen, solid] table [x={t}, y={r}] {\dat};
    
    \pgfplotstableread{data/open_loop/10.dat}{\dat}
    \addplot [ybar,bar width=2pt, mark=none, MyGreen, solid] table [x={t}, y={r}] {\dat};
    
    \pgfplotstableread{data/closed_loop/004/1.dat}{\dat}
    \addplot [ybar,bar width=2pt, mark=none, MyBlue, solid,opacity = 0.3,fill opacity = 0.3] table [x={t}, y={r}] {\dat};
        \pgfplotstableread{data/closed_loop/004/2.dat}{\dat}
    \addplot [ybar,bar width=2pt, mark=none, MyBlue, solid,opacity = 0.3,fill opacity = 0.3] table [x={t}, y={r}] {\dat};
        \pgfplotstableread{data/closed_loop/004/3.dat}{\dat}
    \addplot [ybar,bar width=2pt, mark=none, MyBlue, solid,opacity = 0.3,fill opacity = 0.3] table [x={t}, y={r}] {\dat};
        \pgfplotstableread{data/closed_loop/004/4.dat}{\dat}
    \addplot [ybar,bar width=2pt, mark=none, MyBlue, solid,opacity = 0.3,fill opacity = 0.3] table [x={t}, y={r}] {\dat};
        \pgfplotstableread{data/closed_loop/004/5.dat}{\dat}
    \addplot [ybar,bar width=2pt, mark=none, MyBlue, solid,opacity = 0.3,fill opacity = 0.3] table [x={t}, y={r}] {\dat};
        \pgfplotstableread{data/closed_loop/004/6.dat}{\dat}
   \addplot [ybar,bar width=2pt, mark=none, MyBlue, solid,opacity = 0.3,fill opacity = 0.3] table [x={t}, y={r}] {\dat};
        \pgfplotstableread{data/closed_loop/004/7.dat}{\dat}
   \addplot [ybar,bar width=2pt, mark=none, MyBlue, solid,opacity = 0.3,fill opacity = 0.3] table [x={t}, y={r}] {\dat};
        \pgfplotstableread{data/closed_loop/004/8.dat}{\dat}
    \addplot [ybar,bar width=2pt, mark=none, MyBlue, solid,opacity = 0.3,fill opacity = 0.3] table [x={t}, y={r}] {\dat};
        \pgfplotstableread{data/closed_loop/004/9.dat}{\dat}
    \addplot [ybar,bar width=2pt, mark=none, MyBlue, solid,opacity = 0.3,fill opacity = 0.3] table [x={t}, y={r}] {\dat};
        \pgfplotstableread{data/closed_loop/004/10.dat}{\dat}
    \addplot [ybar,bar width=2pt, mark=none, MyBlue, solid,opacity = 0.3,fill opacity = 0.3] table [x={t}, y={r}] {\dat};
    
    \end{axis}
    \end{tikzpicture}
    \caption{\change{Comparison: open-loop vs closed-loop robust self triggered MPC
    }}
    \label{fig:compare_resource}
\end{figure}

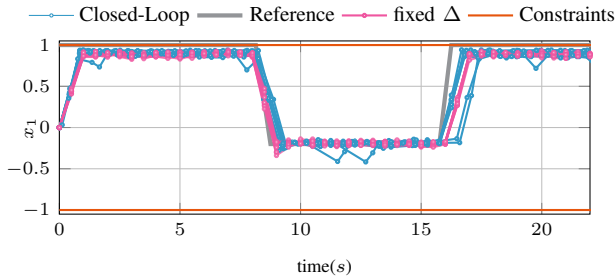
\begin{figure}[htbp!]
    \centering
    \begin{tikzpicture}
    \begin{axis}[xmin=0, xmax=22,
    ymin=-1.05, ymax= 1.05,
    enlargelimits=false,
    clip=true,
    grid=major,
    mark size=0.5pt,
    width=1\linewidth,
    height=0.45\linewidth,ylabel = $x_1$,xlabel= time($s$),
    legend style={
    	font=\footnotesize,
    	draw=none,
		at={(0.5,1.03)},
        anchor=south
    },
    legend columns=4,
    label style={font=\scriptsize},
    ylabel style={at={(axis description cs:0.12,0.5)}},
    xlabel style={at={(axis description cs:0.5,0.02)}},
    ticklabel style = {font=\scriptsize}]
    
    \pgfplotstableread{data/closed_loop/04/1.dat}{\dat}
    \addplot+ [thin, mark=*, mark options={fill=white, scale=1.2},MyBlue] table [x={t}, y={y}] {\dat};
    \addlegendentry{Closed-Loop}
    \addplot+ [ultra thick,Gray,mark=none] table [x={t}, y={ref}] {\dat};
    \addlegendentry{Reference} 
    \addlegendimage{line legend,thick, mark=*, mark options={fill=white, scale=1.2}, VioletRed}
    \addlegendentry{fixed $\Delta$}
    \addplot+ [thick, MyRed, mark = none] table [x={t}, y={max}] {\dat};
    \addplot+ [thick, MyRed, mark = none] table [x={t}, y={min}] {\dat};
    \addlegendentry{Constraints}
    
    \pgfplotstableread{data/closed_loop/04/2.dat}{\dat}
    \addplot [thick, mark=*, mark options={fill=white, scale=1.2}, MyBlue] table [x={t}, y={y}] {\dat};
    
    \pgfplotstableread{data/closed_loop/04/3.dat}{\dat}
    \addplot [thick, mark=*, mark options={fill=white, scale=1.2}, MyBlue] table [x={t}, y={y}] {\dat};
    
    \pgfplotstableread{data/closed_loop/04/4.dat}{\dat}
    \addplot [thick, mark=*, mark options={fill=white, scale=1.2}, MyBlue] table [x={t}, y={y}] {\dat};
    
    \pgfplotstableread{data/closed_loop/04/5.dat}{\dat}
    \addplot [thick, mark=*, mark options={fill=white, scale=1.2},MyBlue] table [x={t}, y={y}] {\dat};
    
    \pgfplotstableread{data/closed_loop/04/6.dat}{\dat}
    \addplot [thick, mark=*, mark options={fill=white, scale=1.2},MyBlue] table [x={t}, y={y}] {\dat};
    
    \pgfplotstableread{data/closed_loop/04/7.dat}{\dat}
    \addplot [thick, mark=*, mark options={fill=white, scale=1.2},MyBlue] table [x={t}, y={y}] {\dat};
    
    \pgfplotstableread{data/closed_loop/04/8.dat}{\dat}
    \addplot [thick, mark=*, mark options={fill=white, scale=1.2},MyBlue] table [x={t}, y={y}] {\dat};
    
    \pgfplotstableread{data/closed_loop/04/9.dat}{\dat}
    \addplot [thick, mark=*, mark options={fill=white, scale=1.2},MyBlue] table [x={t}, y={y}] {\dat};
    
    \pgfplotstableread{data/closed_loop/04/10.dat}{\dat}
    \addplot [thick, mark=*, mark options={fill=white, scale=1.2},MyBlue] table [x={t}, y={y}] {\dat};

    \pgfplotstableread{data/fixed_dt/04/1.dat}{\dat}
    \addplot [thick,opacity = 0.8, mark=*, mark options={fill=white, scale=1.2}, VioletRed] table [x={t}, y={y}] {\dat};
        \pgfplotstableread{data/fixed_dt/04/2.dat}{\dat}
    \addplot [thick,opacity = 0.8,mark=*, mark options={fill=white, scale=1.2}, VioletRed] table [x={t}, y={y}] {\dat};
        \pgfplotstableread{data/fixed_dt/04/3.dat}{\dat}
    \addplot [thick,opacity = 0.8, mark=*, mark options={fill=white, scale=1.2}, VioletRed] table [x={t}, y={y}] {\dat};
        \pgfplotstableread{data/fixed_dt/04/4.dat}{\dat}
    \addplot [thick,opacity = 0.8, mark=*, mark options={fill=white, scale=1.2}, VioletRed] table [x={t}, y={y}] {\dat};
        \pgfplotstableread{data/fixed_dt/04/5.dat}{\dat}
    \addplot [thick,opacity = 0.8,mark=*, mark options={fill=white, scale=1.2}, VioletRed] table [x={t}, y={y}] {\dat};
        \pgfplotstableread{data/fixed_dt/04/6.dat}{\dat}
    \addplot [thick,opacity = 0.8, mark=*, mark options={fill=white, scale=1.2}, VioletRed] table [x={t}, y={y}] {\dat};
        \pgfplotstableread{data/fixed_dt/04/7.dat}{\dat}
    \addplot [thick,opacity = 0.8, mark=*, mark options={fill=white, scale=1.2}, VioletRed] table [x={t}, y={y}] {\dat};
        \pgfplotstableread{data/fixed_dt/04/8.dat}{\dat}
    \addplot [thick,opacity = 0.8, mark=*, mark options={fill=white, scale=1.2}, VioletRed] table [x={t}, y={y}] {\dat};
        \pgfplotstableread{data/fixed_dt/04/9.dat}{\dat}
    \addplot [thick,opacity = 0.8, mark=*, mark options={fill=white, scale=1.2}, VioletRed] table [x={t}, y={y}] {\dat};
        \pgfplotstableread{data/fixed_dt/04/10.dat}{\dat}
    \addplot [thick,opacity = 0.8, mark=*, mark options={fill=white, scale=1.2}, VioletRed] table [x={t}, y={y}] {\dat};
    \end{axis}
    \end{tikzpicture}
    \caption{\change{Output of the proposed controller with larger disturbance}}
    \label{fig:danger_output}
\end{figure}
\begin{figure}[htbp!]
    \centering
    \begin{tikzpicture}
    \begin{axis}[xmin=0, xmax=22,
    ymin= 0 , ymax= 1.05,
    enlargelimits=false,
    clip=true,
    grid=major,
    mark size=0.5pt,
    width=1\linewidth,
    height=0.39\linewidth,ylabel = $r$,xlabel= time($s$),
    legend style={
    	font=\footnotesize,
    	draw=none,
		at={(0.5,1.03)},
        anchor=south
    },
    legend columns=3,
    label style={font=\scriptsize},
    ylabel style={at={(axis description cs:0.06,0.5)}},
    xlabel style={at={(axis description cs:0.5,0.02)}},
    ticklabel style = {font=\scriptsize}]
    
    \pgfplotstableread{data/closed_loop/04/1.dat}{\dat}
    \addplot [ybar,bar width=2pt, mark=none, MyBlue, solid,opacity = 0.3,fill opacity = 0.3] table [x={t}, y={r}] {\dat};
    \addlegendentry{Closed-Loop resource}
    \addplot+ [thick, MyRed, mark = none] table [x={t}, y={rmax}] {\dat};
    \addplot+ [thick, MyRed, mark = none] table [x={t}, y={rmin}] {\dat};
    \addlegendentry{Resource Constraints}
    
        \pgfplotstableread{data/closed_loop/04/2.dat}{\dat}
    \addplot [ybar,bar width=2pt, mark=none, MyBlue, solid,opacity = 0.3,fill opacity = 0.3] table [x={t}, y={r}] {\dat};
        \pgfplotstableread{data/closed_loop/04/3.dat}{\dat}
    \addplot [ybar,bar width=2pt, mark=none, MyBlue, solid,opacity = 0.3,fill opacity = 0.3] table [x={t}, y={r}] {\dat};
        \pgfplotstableread{data/closed_loop/04/4.dat}{\dat}
    \addplot [ybar,bar width=2pt, mark=none, MyBlue, solid,opacity = 0.3,fill opacity = 0.3] table [x={t}, y={r}] {\dat};
        \pgfplotstableread{data/closed_loop/04/5.dat}{\dat}
    \addplot [ybar,bar width=2pt, mark=none, MyBlue, solid,opacity = 0.3,fill opacity = 0.3] table [x={t}, y={r}] {\dat};
        \pgfplotstableread{data/closed_loop/04/6.dat}{\dat}
   \addplot [ybar,bar width=2pt, mark=none, MyBlue, solid,opacity = 0.3,fill opacity = 0.3] table [x={t}, y={r}] {\dat};
        \pgfplotstableread{data/closed_loop/04/7.dat}{\dat}
   \addplot [ybar,bar width=2pt, mark=none, MyBlue, solid,opacity = 0.3,fill opacity = 0.3] table [x={t}, y={r}] {\dat};
        \pgfplotstableread{data/closed_loop/04/8.dat}{\dat}
    \addplot [ybar,bar width=2pt, mark=none, MyBlue, solid,opacity = 0.3,fill opacity = 0.3] table [x={t}, y={r}] {\dat};
    \addplot [ybar,bar width=2pt, mark=none, MyBlue, solid,opacity = 0.3,fill opacity = 0.3] table [x={t}, y={r}] {\dat};
        \pgfplotstableread{data/closed_loop/04/10.dat}{\dat}
    \addplot [ybar,bar width=2pt, mark=none, MyBlue, solid,opacity = 0.3,fill opacity = 0.3] table [x={t}, y={r}] {\dat};
    
    \end{axis}
    \end{tikzpicture}
    \caption{\change{Resource response of the proposed controller with larger disturbance}}
    \label{fig:resource_danger}
\end{figure}
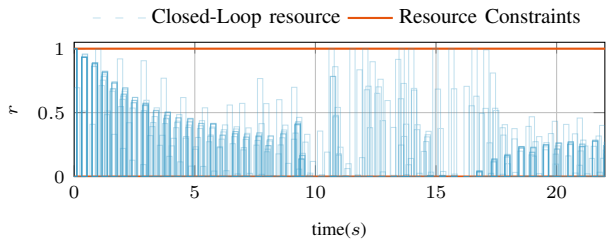

\change{To show the advantage of the proposed scheme over a controller with a fixed sample period $\Delta$, a second experiment has been conducted with a disturbance 10 times stronger, $w(t)\in[-0.4,0.4]$, whose output and resource responses are shown in Figure~\ref{fig:danger_output} and~\ref{fig:resource_danger}, respectively. In both cases, the proposed controller shows comparable performance against the fixed $\Delta$ controller. However, the proposed controller triggers less frequently, with an adaptivity to the working condition, such that the average $\Delta$ is $0.82s$ and $0.492s$ in these two cases, respectively. Figure~\ref{fig:resource_danger} also shows this adaptivity in the resource consumption: from $0s$ to around $9s$, the controller triggers slightly faster than $0.5s$ by consuming the initial resource. Meanwhile, when the reference is far from the output constraint at round $10s$ to $16s$, the proposed controller triggers slightly faster than $0.5s$, which is visualized by the recharging pattern from around $9s$ to $16s$ in Figure~\ref{fig:resource_danger}}. 

\section{Conclusion}\label{sect:conclusion}
This paper proposes a novel resource-aware robust self-triggered MPC, which generalizes resource-aware self-triggered MPC to an uncertain environment. The dynamics ellipsoidal outer approximation of the reachable sets that are governed by a discrete-time feedback control law, is derived to accommodate a continuous-time uncertain disturbance. This feedback law is intentionally designed to be compatible with a self-triggered control scheme. Finally, the proposed scheme is validated through a numerical example.

\bibliographystyle{ieeetr}
\bibliography{ref.bib}

\begin{thebibliography}{10}

\bibitem{heemels2012introduction}
W.~Heemels, K.~Johansson, and P.~Tabuada, ``An introduction to event-triggered
  and self-triggered control,'' in {\em Proc. 51st IEEE Conf. Decis. Control
  (CDC)}, pp.~3270--3285, 2012.

\bibitem{wildhagen2020resource}
S.~Wildhagen, C.~N. Jones, and F.~Allg{\"o}wer, ``A resource-aware approach to
  self-triggered model predictive control,'' {\em IFAC-PapersOnLine}, vol.~53,
  no.~2, pp.~2733--2738, 2020.

\bibitem{Lian2020}
Y.~Lian, S.~Wildhagen, Y.~Jiang, B.~Houska, F.~Allgöwer, and C.~N. Jones,
  ``Resource-aware asynchronous multi-agent coordination via self-triggered
  mpc,'' in {\em Proc. 59th IEEE Conf. Decis. Control (CDC)}, pp.~685--690,
  2020.

\bibitem{Lian2021}
Y.~Lian, Y.~Jiang, N.~Stricker, L.~Thiele, and C.~N. Jones, ``Resource-aware
  stochastic self-triggered model predictive control,'' {\em IEEE Control Syst.
  Lett.}, vol.~6, pp.~1262--1267, 2022.

\bibitem{stricker2021distributed}
N.~Stricker, Y.~Lian, Y.~Jiang, C.~Jones, and L.~Thiele, ``Joint energy
  management for distributed energy harvesting systems,'' in {\em Proceedings
  of The 19th ACM Conference on Embedded Networked Sensor Systems (SenSys'21),
  November 15--17, 2021, Coimbra, Portugal}, (New York, NY, USA), ACM, 2021.

\bibitem{aydiner2015robust}
E.~Aydiner, F.~D. Brunner, W.~P. M.~H. Heemels, and F.~Allgö~wer, ``Robust
  self-triggered model predictive control for constrained discrete-time lti
  systems based on homothetic tubes,'' in {\em Proc. 14th Eur. Control Conf.
  (ECC)}, pp.~1587--1593, 2015.

\bibitem{brunner2016robust}
F.~D. Brunner, M.~Heemels, and F.~Allg{\"o}wer, ``Robust self-triggered mpc for
  constrained linear systems: A tube-based approach,'' {\em Automatica},
  vol.~72, pp.~73--83, 2016.

\bibitem{li2014event}
H.~Li and Y.~Shi, ``Event-triggered robust model predictive control of
  continuous-time nonlinear systems,'' {\em Automatica}, vol.~50, no.~5,
  pp.~1507--1513, 2014.

\bibitem{dai2020fast}
L.~Dai, M.~Cannon, F.~Yang, and S.~Yan, ``Fast self-triggered mpc for
  constrained linear systems with additive disturbances,'' {\em IEEE Trans.
  Autom. Control}, vol.~66, no.~8, pp.~3624--3637, 2021.

\bibitem{farina2012tube}
M.~Farina and R.~Scattolini, ``Tube-based robust sampled-data mpc for linear
  continuous-time systems,'' {\em Automatica}, vol.~48, no.~7, pp.~1473--1476,
  2012.

\bibitem{bock1984multiple}
H.~G. Bock and K.-J. Plitt, ``A multiple shooting algorithm for direct solution
  of optimal control problems,'' {\em IFAC-PapersOnLine}, vol.~17, no.~2,
  pp.~1603--1608, 1984.

\bibitem{borrelli2017predictive}
F.~Borrelli, A.~Bemporad, and M.~Morari, {\em Predictive control for linear and
  hybrid systems}.
\newblock Cambridge University Press, 2017.

\bibitem{villanueva2015unified}
M.~E. Villanueva, B.~Houska, and B.~Chachuat, ``Unified framework for the
  propagation of continuous-time enclosures for parametric nonlinear odes,''
  {\em J. Global Optim}, vol.~62, no.~3, pp.~575--613, 2015.

\bibitem{kurzhanskiui1997ellipsoidal}
A.~Kurzhanskiui and I.~V{\'a}lyi, {\em Ellipsoidal calculus for estimation and
  control}.
\newblock {Nelson Thornes}, 1992.

\bibitem{Houska2019}
B.~Houska and M.~Villanueva, ch.~Robust optimization for MPC, p.~415–447.
\newblock Birkh\"auser.

\bibitem{kurzhanski2002reachability}
A.~Kurzhanski and P.~Varaiya, ``Reachability analysis for uncertain systems-the
  ellipsoidal technique,'' {\em Dynamics of Continuous Discrete and Impulsive
  Systems Series B}, vol.~9, pp.~347--368, 2002.

\bibitem{brockman1998quadratic}
M.~L. Brockman and M.~Corless, ``Quadratic boundedness of nominally linear
  systems,'' {\em Int. J. Contro}, vol.~71, no.~6, pp.~1105--1117, 1998.

\bibitem{schweppe1973uncertain}
F.~C. Schweppe, {\em Uncertain dynamic systems}.
\newblock Prentice Hall, 1973.

\bibitem{steele2004cauchy}
J.~M. Steele, {\em The Cauchy-Schwarz master class: an introduction to the art
  of mathematical inequalities}.
\newblock Cambridge University Press, 2004.

\bibitem{levine2018handbook}
W.~S. Levine, L.~Gr{\"u}ne, {\em et~al.}, {\em Handbook of model predictive
  control}.
\newblock Springer, 2018.

\bibitem{houska2011robust}
B.~Houska, {\em Robust optimization of dynamic systems}.
\newblock PhD thesis, PhD thesis, Katholieke Universiteit Leuven, 2011.

\end{thebibliography}

\end{document}